
\input phyzzx.tex
\tolerance=1000
\vsize = 22 true cm
\hsize = 17 true cm
\hoffset = 0 true cm
\doublespace
\def\ick{\eqalignno}

\medskip
{\baselineskip=12 pt
\line{\hfill UdeM-LPN-TH-93-185}
\line{\hfill CRM-1927}
}
\medskip
\centerline {\bf Structure of the Effective Potential in}
\centerline {\bf Nonrelativistic Chern-Simons Field Theory}
\footnote*{This work is supported in part by funds provided by the
Natural Sciences and Engineering
Research Council of Canada and the Fonds pour la Formation de Chercheurs et
l'Aide \`a la Recherche.}

\vskip 30 pt

\centerline {\bf D. Caenepeel$^{(1)}$, F. Gingras$^{(2)}$,
M. Leblanc$^{(1,2)}$, and D.G.C. McKeon$^{(3)}$}

\vskip 20 pt
{\baselineskip=12 pt
\it
\centerline {\bf \it $\ ^{(1)}$Laboratoire de Physique Nucl\'eaire }
\centerline {\bf \it Universit\'e de Montr\'eal }
\centerline {\bf \it Case postale 6128, succ. A}
\centerline {\bf \it Montr\'eal, (Qc), Canada}
\centerline {\bf \it H3C-3J7 } ,}

\vskip 8 pt

{\baselineskip=12 pt
\it
\centerline {\bf \it $\ ^{(2)}$Centre de Recherches Math\'ematiques}
\centerline {\bf \it Universit\'e de Montr\'eal }
\centerline {\bf \it Case postale 6128, succ. A}
\centerline {\bf \it Montr\'eal, (Qc), Canada}
\centerline {\bf \it H3C-3J7 }}
\medskip
\centerline {and}
\vskip 10 pt

{\baselineskip=12 pt
\it
\centerline {\bf \it $\ ^{(3)}$Department of Applied Mathematics }
\centerline {\bf \it University of Western Ontario }
\centerline {\bf \it London, Ontario, Canada }
\centerline {\bf \it N6A 5B9 }}

\vskip 30 pt
\centerline{\bf Submitted to: {\it Physical Review D}}

\vfill\eject

\centerline {\bf ABSTRACT }
\medskip
We present the scalar field effective potential for nonrelativistic
self-interacting scalar and fermion fields coupled to an Abelian Chern-Simons
gauge field. Fermions are non-minimally coupled to the gauge field via a Pauli
interaction. Gauss's law linearly relates the magnetic field to the matter
field densities; hence, we also include radiative effects from the background
gauge field. However, the scalar field effective potential is transparent to
the presence of the background gauge field to leading order in the
perturbative expansion. We compute the scalar field effective potential in
two gauge families. We perform the calculation in a gauge reminiscent of the
$R_\xi$-gauge in the limit $\xi\rightarrow 0$ and in the Coulomb family
gauges. The scalar field effective potential is the same in both gauge-fixings
and is independent of the gauge-fixing parameter in the Coulomb family gauge.
The conformal symmetry is spontaneously broken except for two values of the
coupling constant, one of which is the self-dual value.
To leading order in the perturbative expansion, the structure of the classical
potential is deeply distorted by radiative corrections and shows a stable
minimum around the origin, which could be of interest when searching for
vortex solutions. We regularize the theory with operator regularization and a
cutoff to demonstrate that the results are independent of the regularization
scheme.

\vfill\eject

\medskip
\noindent {\bf I. Introduction}
\medskip

Chern-Simons theories have been studied in many context in the
last decade from the study of general relativity to condensed matter systems.
An important line of developments occurred when it was shown that classical
relativistic charged scalars minimally coupled to an Abelian Chern-Simons
gauge field in (2+1) spacetime dimensions have vortex
(soliton) solutions
for self-dual equations when the coupling constant takes special values
in a $\phi^6$-theory [1,2]. The presence of vortex solutions permits the
emergence of new mechanisms for anyons superconductivity [3].
Evidence has been found showing that the
existence of such systems possessing vortex solutions is due to
the presence of an $N=2$ supersymmetry obtained by
adding fermion fields in an appropriate way [4,5].

It is more reasonable
to think that the physics of superconductors should be at lower energies
and described by a nonrelativistic system. It turns out
that the  same statements as above can be made
for the corresponding nonrelativistic
field theory. Specifically, by taking the limit $c\rightarrow\infty$
($c$ being the speed of light), one obtains a
field theory of interacting nonrelativistic
scalar fields minimally coupled to an Abelian Chern-Simons gauge field [6,7].
This theory also contains self-dual vortex (soliton) solutions
when the coupling constant takes a special value [7].
Perhaps more surprisingly in the nonrelativistic case,
the self-duality originates also from $N=2$ supersymmetry [8].

The systematic functional method of
Jackiw is an efficient, concise and useful way to evaluate the
effective potential without having to use a classical background field [9].
The functional evaluation of the effective potential
provides us with an exact expression for the one-loop order correction
by summing all 1PI $n$-point graphs evaluated at zero-momentum.

The goal of this paper is to compute the scalar field effective potential of
the nonrelativistic Chern-Simons matter system and to look for deformation of
the effective potential.
\vfill \eject
We start with the action
$$\ick {
S= \int dt d^2{\bf x} \;\;
&\Bigl \{{\kappa\over 2c} (\partial_t {\bf A}) \times {\bf A} -\kappa
A^0 {\bf \nabla} \times {\bf A}
+i\phi^*(\partial_t +ie A^0 )\phi
+i\psi^*(\partial_t +ie A^0)\psi \cr
& - {1 \over 2 } |{\bf D}\phi|^2
 - {1 \over 2 } |{\bf D}\psi|^2
+ {e\over 2c} B |\psi|^2
-{\lambda_1 \over 4}(|\phi|^2)^2
-\lambda_2|\phi|^2 |\psi|^2
\Bigr \}&(1.1) \cr }
$$
where ${\bf D}={\bf \nabla }-i{e\over c}{\bf A}$ is the covariant derivative
and $B={\bf \nabla}\times {\bf A}$ is the magnetic field.
The action (1.1) represents
a system of self-interacting scalar field and fermionic field
coupled to an Abelian Chern-Simons gauge field. Note that the latter is
non-minimally coupled to the gauge field through the Pauli term.
The action (1.1) is $N=2$ supersymmetric when $\lambda_1= -{2e^2\over c\kappa}$
and $\lambda_2={3\over 4}\lambda_1$ [8]. We use
a vector notation: For instance, in the plane the cross product
is ${\bf V}\times {\bf W} = \epsilon^{ij}V^iW^j$, the curl of a vector is
${\bf \nabla}\times {\bf V}=\epsilon^{ij}\partial_iV^j$, the curl of
a scalar is $({\bf\nabla}\times S)^i=\epsilon^{ij}\partial_jS$ and we shall
introduce the notation $\bigl ({\bf A}\times {\bf {\hat z}}\bigr)^i
=\epsilon^{ij}A^j$.
The notation $x=(t, {\bf x})$ will also be used unless stated otherwise.

To analyze the structure of the scalar field effective potential
of the action (1.1), we proceed with
two regularization methods in conjunction with the functional evaluation of
Jackiw. We use 1) operator regularization (OR) [10,11] and 2) the
cutoff regularization method. The first method was
proposed a few years ago and regularizes the
operators present in the theory instead of modifying the original Lagrangian
by either adding counterterms or massive fields. It is an powerful method
for regularization since it preserves
classical symmetries modulo anomalies and at every stage of the calculation
all ultraviolet divergences are absent
even when the regulator is removed. OR has been applied successfully in many
examples of relativistic field theories such as gauge theories, supersymmetry,
chiral anomalies, curved-spacetime QFTs, quantum gravity, and for higher
loop contibutions in many scalar field theories [see 10 and 11, and
refs. therein]. However,
OR has never been applied to nonrelativistic theories
such as the present case. Let us summarize the method.
The effective potential is given to ${\cal O}(\hbar)$ by
the trace of the logarithm of operators
occurring in the theory. The logarithm is then expressed in the form
$$
\ln H=-\lim_{s\to0}{d \over ds}H^{-s}
\eqno(1.2)
$$
where the operator $H$ is dimensionless in OR once a mass parameter $\mu$
is introduced which plays the role of the renormalization scale.
Next, we use the $\Gamma$-function representation
of an operator,
$$
H^{-s}={1\over \Gamma (s)}\int_0^\infty dt\; t^{s-1}\, e^{-Ht}\quad
\eqno (1.3)
$$
and the trace of the exponential in Eq.~(1.3) is given by the Schwinger
expansion
$$\ick {
{\rm Tr}\, e^{-(H_0+H_I)t}= {\rm Tr} \bigl \{e^{-H_0t} &+ (-t) e^{-H_0t}H_I \cr
& + {(-t)^2\over 2}\int_0^1 du e^{-(1-u)H_0 t} H_I e^{-uH_0 t} H_I +\cdots
\bigr \}
&(1.4) \cr}
$$
upon defining $H=H_0+H_I$.
The $n$-point function is easily extracted from the Schwinger expansion as
each $H_I$ corresponds to a 1-point insertion.

The second method of regularization consists in cutting the region of
integration for the momentum vector. In $D$-dimensional relativistic theories,
the cutoff is put on the $D$-vector $p^2=E^2-{\bf p^2}c^2 < \Lambda^2$, which
preserves Lorentz-transformations. In the nonrelativistic theory, the cutoff
is put only on the $(D-1)$-vector, leaving the energy unconstrained.
The cutoff method regulates ultraviolet divergences in such a way that the
radiative corrections are invariant under rotations when used in
nonrelativistic theories. However, it is not clear that this method preserves
Galilean boosts when the regularization is performed (or at least no proof
exists that the method is Galilean invariant). Recently, the cutoff method has
been used by Lozano [12] and Bergman and Lozano [13] in a recent investigation
of a similar problem. In Lozano's analysis, the ground state energy is
evaluated in the same model (except for the inclusion of fermi fields) only at
the value of classical solution in the thermodynamic limit with the
introduction of a chemical potential, and of an external charge at spatial
infinity to maintain charge neutrality. In the second approach, Bergmann and
Lozano used the Feynman diagrammatic with the cutoff method to evaluate the
scattering of two scalars in the same theory.

We would like to show that in
the context of the effective potential method without an external charge but
including a background gauge field consistent with Gauss's law,
the answer is the same irrespective of the method one uses.
In this way, we infer
that the cutoff method preserves Galilean invariance since OR is a
regularization scheme that preserves all classical symmetries up to anomalies.
We conclude further
that the external charge is unnecessary since we obtain an answer
consistent with Gauss's law. We generalize their results by
providing the one-loop evaluation of the scalar field effective potential in
the $R_\xi$-gauge (in the $\xi\to0$ limit)
including radiative effects with external gauge field
and in the Coulomb-gauge family.
The remainder of the
paper is divided in 3 sections. In the next section, we set up the problem
illustrating clearly what is the procedure suggested by the classical
equation of motion when the scalar field is constant. In the third section, we
calculate the scalar field effective potential in the $R_\xi$-gauge
using both regulators. In the fourth section, we discuss the scale anomaly and
the structure of the effective potential. In the appendix, we derive the same
effective potential in the Coulomb-gauge family and show that the result is
gauge-parameter independent in both methods.

\medskip
\noindent {\bf II. Constant Background Matter Field and Equation of Motions}
\medskip

The coupling constants of the theory of Eq.~(1.1) are $\lambda_1$,
the strength of
the scalar field self-interaction, $\lambda_2$, the fermion-scalar
interaction and $\kappa$, the Chern-Simons coupling
constant, which we take positive for convenience. It is easy to see that these
are the only three coupling constants: making the following substitution
$$\ick {
{\bf A}' = {e \over c}{\bf A};\quad
A'^0 &= e A^0;\quad
\kappa '= {\kappa c \over e^2}&(2.1) \cr}$$
\vfill\eject
{\noindent in} the classical action (1.1)
and dropping the primes gives
$$\ick {
S=\int dt d^2{\bf x}\;\Bigl \{
&{\kappa\over 2} (\partial_t {\bf A})\times {\bf A}-\kappa
A^0 {\bf\nabla}\times {\bf A}
-A^0(|\phi|^2 +|\psi|^2 )\cr &
-{i\over 2} {\bf A}\cdot
\Bigl(\phi^* \nabla\phi - (\nabla\phi^*)\phi +
\psi^* \nabla\psi - (\nabla\psi^*)\psi \Bigr)
-{1\over 2} {\bf A}\cdot {\bf A} (|\phi|^2 + |\psi|^2 ) + {1\over 2}
{\bf \nabla}\times {\bf A} |\psi|^2 \cr
&+\phi^*(i\partial_t +{1 \over 2} {\bf \nabla^2})\phi
+\psi^*(i\partial_t +{1 \over 2} {\bf \nabla^2})\psi
- {\lambda_1 \over 4} (|\phi|^2)^2 -\lambda_2 |\phi|^2 |\psi|^2 \Bigr \}
&(2.2) \cr}
$$
where an integration by parts has been performed.

The effective potential method starts with the definition of a new
shifted action:
$$\ick {
S_{\rm new}&=S\Bigl\{\phi(x)=\varphi+\pi(x);\psi(x);
A^\mu (x)=a^\mu(x)+Q^\mu(x)\Bigr \}\cr
&\quad -S\Bigl \{ \varphi,a^\mu(x)\Bigr\} - {\rm terms\;linear \;in \;quantum\;
fields,} &(2.3)\cr }
$$
where we shift the scalar field by a constant field and we shift
the gauge field by a solution to the classical equations of motion for the
electromagnetic fields. The
fermionic field is not shifted because we consider only quantum corrections
to the scalar field effective potential. The important question is to know
how to choose
the background gauge field $a^\mu(x)$. For this purpose, we need to analyse
the classical equations of motion with constant scalar field
configuration. The classical equations of motion for arbitrary
scalar field and arbitrary gauge field (with vanishing classical
fermion field) are
$$\ick {
B&= {\bf \nabla}\times {\bf a}=-{1\over \kappa} |\phi|^2 &(2.4a) \cr
{\bf E} &\equiv - {\bf \nabla} a^0 - \partial_t {\bf a}
= {1\over \kappa} {\bf J}\times {\bf {\hat z}}
&(2.4b) \cr
\Bigl [ i(\partial_t +i&a^0) + {1\over 2} {\bf D}^2(a^\mu) - {\lambda_1\over 2}
|\phi|^2 \Bigr ] \phi = 0 &(2.4c) \cr }
$$
where the current is given by
${\bf J} ={1\over 2i}
\bigl [
\phi^* {\bf D}(a^\mu)\phi - ({\bf D}(a^\mu)\phi)^*\phi \bigr ] $
and ${\bf D}(a^\mu)$ is
the covariant derivative with respect to the background gauge field.
The equation for the
magnetic field (2.4a) is recognized as Gauss's law.  In order to apply the
functional method of Jackiw, the scalar field needs to be shifted
by a constant and Eqs.(2.4) are read with $\phi=\varphi={\rm constant}$.
To maintain consistency with Gauss's law, we need to
choose a background gauge field $a^\mu(x)$ such that the magnetic field is
constant throughout the plane. We set ${\bf a}({\bf x})
= -{B\over 2}{\bf x}\times {\bf {\hat z}}
={\varphi^*\varphi\over 2\kappa}{\bf x}\times {\bf {\hat z}}$ where $B$ is the
constant magnetic field. Such a choice is also
consistent with the electric field equation of motion if
$a^0({\bf x})=-{(\varphi^*\varphi)^2\over 4\kappa^2}
{\bf x}^2$.
The above choice for $a^\mu({\bf x})$
provides a solution for the equation of motion
for the electromagnetic field when the system has constant scalar field
configuration. However, the equation of motion for the scalar field Eq.~(2.4c)
is not satisfied with constant scalar field and with the above gauge field
choice unless $\varphi=0$.
This is not too surprising since the extremum of our classical potential
is located at the origin.

The advantage of the present setup for calculating radiative corrections
to the scalar field
effective potential is that we do not need to introduce an external
charge at spatial infinity, nor do we need to introduce a
chemical potential thus of an unnatural classical symmetry-breaking solution
away from the origin [12]. However, we need to consider the background gauge
field $a^\mu({\bf x})$ because it could lead $\varphi$-dependence for the
scalar field effective potential.
We now turn to the calculation of the effective potential in the
$R_\xi$-gauge using the functional method in conjunction with OR.

\medskip
\noindent {\bf III. Effective Potential in the $R_\xi$-gauge. }
\medskip

In order to quantize the theory, we need to gauge-fix the action (2.3) on the
quantum gauge field. A simplification in the calculation occurs when the
gauge-fixing term
$${\cal L}_{G.F.}={1\over 2\xi}\bigl [ {\bf \nabla }\cdot{\bf Q}
+ i\xi\varphi\pi^*\bigr]
\bigr [ {\bf \nabla }\cdot {\bf Q} - i\xi\varphi^*\pi\bigr]$$
reminiscent of the $R_\xi$-gauge is chosen (the Coulomb family gauges
${\cal L}_{G.F.}= {1\over 2\xi}({\bf \nabla}\cdot {\bf Q})^2$
will be treated in the appendix).
The choice of the $R_\xi$-gauge enables us to eliminate some of
the cross-terms mixing gauge and scalar fields. The quadratic action in the
quantum fields then
\vfill\eject
{\noindent becomes }
$$\ick {
S=\int dt \; d^2{\bf x} \; &
\Bigl\{ {\kappa\over 2}(\partial_t{\bf Q})\times {\bf Q}
-\kappa Q^0 {\bf \nabla}\times {\bf Q}
+ {1\over 2\xi}({\bf \nabla}\cdot {\bf Q})^2 - {\rho\over 2}{\bf Q}\cdot
{\bf Q}\cr
&+i\pi^*(\partial_t +ia^0)\pi -{1\over 2}|{\bf D}\pi|^2 - {\lambda_1\over 4}
\Bigl (\varphi^2 (\pi^*)^2+ 4 \rho |\pi|^2 + (\varphi^*)^2(\pi)^2
\Bigr )+{\xi\over 2}\rho |\pi|^2 \cr
&+i\psi^*(\partial_t +ia^0)\psi -{1\over 2}|{\bf D}\psi|^2 -
\lambda_2 \rho |\psi|^2
+{B \over 2}|\psi|^2 \cr
&+c^*(-\nabla^2 +\xi\rho )c
+J^0Q^0 + {\bf J}\cdot {\bf Q}\Bigr\} &(3.1)\cr }
$$
where the current has the form $J^0=-[\varphi^*\pi+\pi^*\varphi]$,
${\bf J}={\bf a}
J^0$ and the notation $\rho=\varphi^*\varphi$ is used. The $c$-field term in
Eq.(3.1) is the ghost compensating term arising from
the choice of gauge-fixing condition. Thus the quadratic part in the spacetime
varying field of the shifted action (3.1) in the $R_\xi$-gauge has the form
$$\ick {
\int dt d^2{\bf x}\;{\cal L}'(\varphi, a^\mu({\bf x}),
&\pi^a(x), \psi^a(x), c(x),
Q^\mu(x))= \int dt dt' d^2{\bf x} d^2{\bf x'} \;\Bigl \{
{1\over 2} \pi^{*a}(x) {\cal D}^{-1}_{ab}(x-x') \pi^b(x')\cr
&+{1\over 2} \psi^{*a}(x) {\cal S}^{-1}_{ab}(x-x') \psi^b(x')
-{1\over 2}Q^\mu(x) {\Delta}^{-1}_{\mu\nu}(x-x')Q^\nu (x')\cr
&+c^*(x){\cal P}^{-1}(x-x')c(x')
+ {1\over 2} J^\mu (x) {\Delta}^{\mu\nu} (x-x') J^\nu (x') \Bigr \}
&(3.2) \cr}
$$
where the notation for the scalar and fermion fields is $\pi^a = (\pi, \pi^*)$;
$\psi^a = (\psi, \psi^*)$; $a=1,2$; $Q^\mu = ({Q^0\over c}, Q^1, Q^2)$. The
last term involving currents is obtained by a conventional shift of variable
involving the quantum gauge field (in contrast to the procedure used in the
appendix). The matrices ${\cal D}^{-1}$, ${\cal S}^{-1}$, ${\Delta}^{-1}$ and
${\cal P}^{-1}$ can be read from the action (3.1) and will be used below.

We now discuss the structure of the perturbative expansion.
The effective potential is related to the effective action by
$V_{\rm eff}\int d^3x=-\Gamma_{\rm eff}$ when
defined on constant background fields. In the present case, the background
gauge field $a^\mu({\bf x})$
is space-dependent; hence, we cannot use directly the functional
method of Jackiw.
We adopt the following strategy. We will compute the effective action by
factoring out a matrix that is
background gauge field independent and perturbatively expand the gauge field
dependent part in powers of small coupling constants $\lambda_1\ll 1$,
$\kappa^{-1}\ll 1$ (recall that $a^0\sim {\rho^2\over
\kappa^2}$ and ${\bf a }\sim {\rho\over \kappa}$).
The computation is up to ${\cal O}(\rho^3)$ because each term of
${\cal O}(\rho^3)$ is either of ${\cal O}(\lambda_1^3)$,
${\cal O}({\lambda_1^2\over \kappa})$ or ${\cal O}({1\over \kappa^3})$.
Therefore, for the rest of the paper, we will use the terminology
${\cal O}(\rho^3)$ to mean that the expansion is in small coupling
constants to ${\cal O}(\lambda_1^3)$,
${\cal O}({\lambda_1^2\over \kappa})$ or ${\cal O}({1\over \kappa^3})$.
We do not introduce the parameter $\lambda_2$ in these expressions for
simplicity since
$\lambda_2$ does not enter the scalar field effective potential, as we
will see below. Thus, as again we will see, we need only to consider 1-point or
2-point functions in background gauge fields that contribute to the effective
action. The gauge-independent part will be treated following Jackiw's
method [9].

{}From the fact that the Lagrangian is quadratic in quantum fields,
we can easily perform the
Gaussian integrals. The effective action to
${\cal O}(\hbar)$ is given as usual by
$$
\Gamma_{\rm eff}= S(\varphi,a^\mu({\bf x})) +
                  {i\over 2}\ln {\rm Det} \{{\cal D}^{-1}_{ab} + {\cal
M}_{ab}\}
-i \ln {\rm Det} {\cal S}^{-1}_{ab}
+{i\over 2} \ln {\rm Det} \Delta^{-1}_{\mu\nu} - i \ln {\rm Det} {\cal P}^{-1}
\eqno (3.3)
$$
where the determinant ${\rm Det}$ is functional.
The matrices are defined at the
operator level where $\omega$ and ${\bf p}$ are operators acting on functions
on
the right as follows:
$$
{\cal D}^{-1}(\varphi, a^\mu ({\bf x});\omega, {\bf p})
= \pmatrix { \omega +{1\over 2}{\bf D}\cdot{\bf D}
+({\xi\over 2}-\lambda_1)\rho
-a^0
&-{\lambda_1\over 2}\varphi\varphi \cr
-{\lambda_1\over 2}\varphi^*\varphi^*
&-\omega+{1\over 2}{\bf D}^*\cdot{\bf D}^* +({\xi\over 2}-\lambda_1)
\rho-a^0\cr
}\eqno (3.4)
$$
where ${\bf D}=i({\bf p} - {\bf a}({\bf x}))$. The matrix
${\cal M}$ is obtained from the current-current term of Eq.(3.2)
$$
{\cal M}(\varphi, a^\mu ({\bf x});\omega, {\bf p})
= {\bigl [-\rho +i\kappa ({\bf a}\times {\bf p}-
{\bf p}\times {\bf a}) +{\cal O}(\xi)
\bigr]
\over \kappa ({\bf p}^2-\xi\rho)}
\pmatrix { \rho &\varphi\varphi \cr
\varphi^*\varphi^*
&\rho \cr
}\quad ,\eqno (3.5)
$$
and the matrix for fermions is
$$
{\cal S}^{-1}(\varphi, a^\mu ({\bf x});\omega, {\bf p})
= \pmatrix { \omega +{1\over 2}{\bf D}\cdot{\bf D}
-\lambda_2\rho + {1\over 2}B-a^0
&0 \cr
0
&\omega-{1\over 2}{\bf D}^*\cdot{\bf D}^*+\lambda_2\rho-{1\over 2}B+a^0\cr
}\quad .\eqno (3.6)
$$
The gauge field matrix is
$$
{\Delta}^{-1}(\varphi;\omega, {\bf p})=
\pmatrix {0&-ic\kappa p^2 &ic\kappa p^1\cr
          ic\kappa p^2 &\rho-{1\over \xi}p^1p^1 &-i\kappa\omega-{1\over\xi}
p^1p^2\cr
          -ic\kappa p^1 &i\kappa\omega-{1\over \xi}p^1p^2 &
\rho-{1\over \xi}p^2p^2 \cr }\quad ,\eqno (3.7)
$$
and the ghost matrix is
$$
{\cal P}^{-1}(\varphi;\omega,{\bf p})=({\bf p}^2 +\xi\rho)\quad .\eqno (3.8)
$$

Now, let us use the limit $\xi\to 0$. In this limit, contributions to the
effective action
from integrating the quantum gauge fields and
ghosts disappear since the $\ln\det \Delta^{-1}$ and the
$\ln\det {\cal P}^{-1}$ become $\rho$-independent.
The remaining contributions are from the
matter fields. We transform the $\ln {\rm Det}$ as a functional trace and
logarithm, $\ln {\rm Det}= {\rm Tr} {\rm Ln} $. Then, we separate a background
gauge field dependent matrix as ${\cal D}^{-1}+{\cal M}= \Theta^{-1} \bigl ( 1
+ \Theta X\bigr )$
and similarly for the fermions
${\cal S}^{-1}= \Omega^{-1}\bigl ( 1+\Omega Y\bigr)$
where the matrices $\Theta^{-1}$ and $\Omega^{-1}$ are
background gauge field independent. Then, for example,
for the scalar field
$$\ick{
{\rm Tr Ln} ( {\cal D}^{-1} +{\cal M})&= {\rm Tr Ln} \Theta^{-1}
+ {\rm Tr Ln}\bigl ( 1 + \Theta X\bigr )\cr
&={\rm tr} \ln \det \Theta^{-1} + {\rm Tr Ln}
\bigl ( 1 + \Theta X\bigr )&(3.9)\cr}
$$
where for the first term of the final line, the
trace is on momentum/energy  space and the determinant
on internal space
indices. The last term of Eq.(3.9) is the background gauge field
dependent part and we find its contribution to the effective action
by a perturbative expansion. A similar expression for the fermions
also arises. It is
easy to find the matrices $\Theta^{-1}$, $\Theta$,
$\Omega^{-1}$, $\Omega$, $X$, and $Y$ to the order needed. They are
$$
\Theta^{-1}
= \pmatrix { \omega -{1\over 2}{\bf p^2}-\lambda_1\rho -{\rho^2\over \kappa^2
{\bf p^2}}
&(-{\lambda_1\over 2}-{\rho\over\kappa^2{\bf p^2}})\varphi\varphi \cr
(-{\lambda_1\over 2}-{\rho\over\kappa^2{\bf p^2}})\varphi^*\varphi^*
&-\omega-{1\over 2}{\bf p^2}-\lambda_1\rho - {\rho^2\over\kappa^2 {\bf p^2}}
}\quad ,\eqno (3.10a)
$$
$$
\Omega^{-1}
= \pmatrix { \omega -{1\over 2}{\bf p^2}-\lambda_2\rho +{1\over 2} B
&0\cr
0
&\omega+{1\over 2}{\bf p^2}+\lambda_2\rho - {1\over 2} B
}\quad ,\eqno (3.10b)
$$
and for their inverse,
$$
\Theta
={1\over \bigl[-\omega^2+({1\over 2}{\bf p}^2
+\lambda_1\rho)^2\bigr]} \Bigl [ \pmatrix { -\omega -{1\over 2}{\bf
p^2}-\lambda_1\rho
&{\lambda_1\over 2}\varphi\varphi  \cr
{\lambda_1\over 2}\varphi^*\varphi^*
&\omega-{1\over 2}{\bf p^2}-\lambda_1\rho }  +
\Theta_1({\cal O} (\rho^2))\Bigr ],
\eqno (3.10c)
$$

{\noindent where} $\Theta_1$ is of ${\cal O}(\rho^2)$ and

$$
\Omega
={1\over \bigl[\omega^2-({1\over 2}{\bf p}^2 +\lambda_2\rho -{1\over 2} B)^2
\bigr]} \pmatrix { \omega +{1\over 2}{\bf p^2}+\lambda_2\rho -{1\over 2} B
&0\cr
0
&\omega-{1\over 2}{\bf p^2}-\lambda_2\rho + {1\over 2} B
}\quad .
\eqno (3.10d)
$$

{\noindent The} matrices
$X$ and $Y$ are the gauge field dependent matrices of a similar form

$$
X
= \pmatrix { {1\over 2} ({\bf a}\cdot {\bf p} +{\bf a}\cdot {\bf p})
&0\cr
0
& -{1\over 2} ({\bf a}\cdot {\bf p} +{\bf a}\cdot {\bf p}) }
+X_1({\cal O}(\rho^2) )\eqno (3.11a)
$$
and
$$
Y
= \pmatrix { {1\over 2} ({\bf a}\cdot {\bf p} +{\bf a}\cdot {\bf p})
&0 \cr
0
& {1\over 2} ({\bf a}\cdot {\bf p} +{\bf a}\cdot {\bf p}) }
+Y_1({\cal O}(\rho^2) )\eqno (3.11b)
$$

{\noindent We} have included the background constant magnetic field in the
fermion $\Omega^{-1}$--matrix for simplicity and the two matrices
$X_1$ and $Y_1$ are background gauge field dependent and of ${\cal O}(\rho^2)$.
Let us first consider the
background gauge field contribution coming from the last term of Eq.~(3.9)
$$
{\rm Tr Ln} (1+\Theta X)={\rm Tr}
[\Theta X-{1\over 2}(\Theta X)(\Theta X) + ...]
\eqno(3.12)
$$
where the matrix $X$ is at least of ${\cal O}(\rho)$. The first term in (3.12)
represents 1-point function contributions
(alternatively called ``tadpole" graphs) with one external gauge field
dependent line. No external momentum enters in those graphs. For example,
one of the tadpole graphs is given by
$$\ick{
&\int d^2{\bf p} d\omega \langle {\bf p}\omega\vert{1\over
\bigl(-\omega^2+({1\over 2}{\bf p}^2 + \lambda_1\rho)^2\bigr)} (-\omega
-{1\over
2}{\bf p}^2-\lambda_1\rho)({\bf a}\cdot {\bf p})\vert\omega {\bf p}\rangle \cr
&=\int d^2{\bf p}  \;\;{\bf p}\cdot
\langle{\bf p}\vert {\bf a}\vert{\bf p}\rangle\int d\omega
{(-\omega -{1\over 2}{\bf p}^2-\lambda_1\rho)\over
[-\omega^2 +({1\over 2}{\bf p}^2+\lambda_1\rho)^2] }
\langle\omega\vert\omega\rangle
&(3.13)}
$$
which vanishes either because of symmetric integration over the momentum space
or the energy space, or is neglected because it
simply gives ${\cal O}(\rho^3)$ contributions. This is in fact true for all
tadpole graphs encountered in this theory.

The ${\cal O}(\rho^2)$ contribution coming from the
second term in Eq.(3.12) is the 2-point function and is given by
$$
\ick{
-{1\over 8}&\int d^2{\bf p} d\omega_p\langle {\bf p}\omega_p\vert {1\over
\bigl(-\omega^2+({1\over 2}{\bf p}^2 + \lambda_1\rho)^2\bigr)}
(\omega +{1\over2}{\bf p}^2)
({\bf a}\cdot{\bf p}+{\bf p}\cdot {\bf a}) \cr \times
&\int d^2{\bf q} d\omega_q\vert {\bf
q}\omega_q\rangle\langle {\bf q}\omega_q\vert
{1\over
\bigl(-\omega^2+({1\over 2}{\bf p}^2 + \lambda_1\rho)^2\bigr)}
(\omega + {1\over2}{\bf p}^2)
({\bf a}\cdot{\bf p}+{\bf p}\cdot {\bf a})\vert{\bf p}\omega_p\rangle \cr
-{1\over 8}&\int d^2{\bf p} d\omega_p\langle {\bf p}\omega_p\vert {1\over
\bigl(-\omega^2+({1\over 2}{\bf p}^2 + \lambda_1\rho)^2\bigr)}
(\omega -{1\over2}{\bf p}^2)
({\bf a}\cdot{\bf p}+{\bf p}\cdot {\bf a})\cr \times
&\int d^2{\bf q} d\omega_q\vert {\bf
q}\omega_q\rangle\langle {\bf q}\omega_q\vert
{1\over
\bigl(-\omega^2+({1\over 2}{\bf p}^2 + \lambda_1\rho)^2\bigr)}
(\omega -{1\over2}{\bf p}^2)
({\bf a}\cdot{\bf p}+{\bf p}\cdot {\bf a})
\vert{\bf p}\omega_p\rangle &(3.14) \cr
}
$$
Upon acting on states, integrating
$\delta(\omega_q-\omega_p)$ which arises because ${\bf a}$ is only
${\bf x}$-dependent, we get
$$\ick {
-{1\over 8}\int d^2{\bf p}&d^2{\bf q} d\omega \Bigl \{
{1\over \bigl(-\omega+({1\over 2}{\bf p}^2 + \lambda_1\rho)\bigr)
\bigl(-\omega+({1\over 2}{\bf q}^2 + \lambda_1\rho)\bigr)} + \cr
& + {1\over \bigl(\omega+({1\over 2}{\bf p}^2 + \lambda_1\rho)\bigr)
\bigl(\omega+({1\over 2}{\bf q}^2 + \lambda_1\rho)\bigr)} \Bigr \}
[{\bf p}+{\bf q}]\cdot {\bf a}({\bf p}-{\bf q})
[{\bf p}+{\bf q}]\cdot {\bf a}({\bf q}-{\bf p})\cr
& \;\;\;
&(3.15) \cr
}
$$

{\noindent Now}, Eq.~(3.15) is easily evaluated if one notices that upon using
the $i\epsilon$-prescription $\pm\omega+{1\over 2}{\bf p}^2 + i\epsilon$ a
contour integration on $\omega$ necessarily gives zero for each integral
since all the poles (for each integrals) in $\omega$ are located on the same
side of the real axis. A similar analysis for the fermions gives the same
result for the background gauge field contribution. We therefore conclude that
to ${\cal O}(\rho^3)$ the background gauge field does not contribute to the
effective action/potential in the $R_\xi$-gauge with $\xi\to 0$; hence, all
contributions come from the first term in Eq.~(3.9), {\it i.e.,}
the background gauge
field independent term.

We evaluate the first term of Eq.~(3.9) and the similar expression coming
from the fermions
$$\ick {
V_{\rm eff} = -{\Gamma_{\rm eff}\over \int d^3x}
=V_0(\rho)
&-{i \over 2}\int {d^2{\bf p} \over (2\pi)^2} {d\omega \over 2\pi}\ln
{1\over \mu'^4}
\Bigl \{ -\omega^2 + \bigl ({{\bf p}^2 \over 2} +\lambda_1\rho\bigr )^2
+ \bigl (
-{\lambda_1^2 \over 4}+ {1 \over \kappa^2} \Bigr )\rho^2
+{\cal O}(\rho^3)
\Bigr \} \cr
&+i \int {d^2{\bf p} \over (2\pi)^2} {d\omega \over 2\pi}\ln
{1\over \mu'^4}
\Bigl \{ -\omega^2 + \bigl ({{\bf p}^2 \over 2} +\lambda_2\rho-{1\over 2}B
\bigr )^2  \Bigr \}
. &(3.16) \cr  }
$$
where we have performed the determinant of the matrix $\Theta^{-1}$ and
$\Omega^{-1}$
and $V_0(\rho)={1 \over 2}\delta m\, \rho+{1 \over 4}(\lambda_1+\delta
\lambda_1)\rho^2$
with $\delta m$ and $\delta\lambda_1$ the counterterms, anticipating
use of the cutoff regularization method.
Up to this point in the evaluation of the effective potential, we did not
use any regularization prescription. Divergences, however, now occur in the
momentum integration. We perform the computation
first in the cutoff and then in the OR method.  In the cutoff method,
both integrals of Eq.~(3.16)
are simplified using $-{i\over 2}\int {d\omega\over 2\pi} \ln (-\omega^2+E^2-
i\epsilon)={1\over 2}E$. The scalar field effective potential then becomes
$$\ick {
V_{\rm eff}=V_0(\rho) &+ {1\over 2}\int {d^2{\bf p}\over (2\pi)^2}
\Bigl [ ({{\bf p}^2\over 2}+\lambda_1\rho)^2 +(-{\lambda_1^2\over 4}
+{1\over \kappa^2})\rho^2 \Bigr ]^{1/2} \cr
& -\int {d^2{\bf p}\over (2\pi)^2}({{\bf p}^2\over 2}
+\lambda_2\rho -{1\over 2}B ) +{\cal O}(\rho^3)
&(3.17) \cr }
$$
The second
integral from fermions leads to a field-independent infinite term
and a contribution
$a_1(\lambda_2\rho-{1\over 2}B)\Lambda^2$, which will be removed upon
renormalization. The first integral can be expanded in powers of coupling
constants and the scalar field effective action before renormalization
becomes
$$
V_{\rm eff} = V_0(\rho) + \bigl (a_1(\lambda_2\rho-{1\over 2}B)+
a_2\lambda_1\rho\bigr )\Lambda^2
+{1\over 32\pi}\bigl (-\lambda_1^2 + {4\over \kappa^2}\bigr )\rho^2\ln
{\Lambda\over 2\lambda_1\rho} +{\cal O}(\rho^3) \quad .
\eqno (3.18)
$$
The renormalization is performed with normalization conditions
$${d^2 \over d \rho^2}V_{\rm eff}|_{\rho=\mu^2}={1 \over 2}\lambda_1(\mu)
\eqno (3.19)
$$
and vanishing mass. We
obtain to one-loop order, in the cutoff method, the renormalized
scalar field effective potential in the $R_\xi$-gauge in the $\xi\to 0$ limit
$$\ick{
V^R_{\rm eff}(\rho)=&
{\lambda_1(\mu)\over 4}\rho^2+{1\over {32\pi}}\bigl (
\lambda_1^2(\mu)-{4\over \kappa^2}\bigr )\rho^2\,
\bigl(\ln {\rho\over\mu^2}-{3\over 2}\bigr)\quad &(3.20)\cr }
$$
with $\lambda_1(\mu)\approx\lambda_1 +{\cal O}(\lambda_1^2)$.

We next use OR
with the following identification. For the first logarithm of Eq.~(3.16),
we use
$H_0=\bigl
[-\omega ^2+ \bigl ({{\bf p}^2 \over 2}+\lambda_1\rho\bigr )^2\bigr ]/\mu'^4$
and $H_I=\Bigl (-{\lambda_1^2 \over 4}+ {1 \over \kappa^2} \Bigr )\rho^2/
\mu'^4 $
and for the second logarithm, we use
$H_0= \Bigl
[-\omega^2 + \bigl ({{\bf p}^2 \over 2} +\lambda_2\rho-{1\over 2}B
\bigr )^2\Bigr ]/\mu'^4 $; however, for the fermions, there is no $H_I$.
Now, we use Eqs.~(1.2-4) and
expand in powers of $\rho$, i.e., compute only up to the 1-point function
in the Schwinger expansion we get
$$\ick {
V_{\rm eff}(\rho)={\lambda_1 \over 4}\rho^2 + a_3 \rho^2
-{i\over 2}\lim_{s\to 0}\int {d^2 {\bf p}\over (2\pi)^2}
{d\omega\over 2\pi} &({d\over ds}){1\over \Gamma (s)}\times &(3.21)\cr
&\Bigl \{ \mu'^{4s}\rho^2 {
(-{1\over 4}\lambda_1^2 + {1\over\kappa^2})\over
[-\omega^2+({{\bf p}^2\over 2}+\lambda_1\rho)^2]^{1+s} } \Gamma (1+s)
+{\cal O}(\rho^3) \Bigr \} \cr }
$$
where an unimportant constant $a_3$ [see below]
arises from the first term independent of $H_I$ in the
Schwinger expansion. The integration over $\omega$ is easy
to perform in the complex plane
using the residue theorem.  The integral we need
is
$$
I\equiv\int_{-\infty}^\infty
{d\omega\over 2\pi}{1\over \lbrack -\omega^2+({{\bf p}^2 \over
2}+a)^2\rbrack^{1+s}}=i{(2s)!\over s!s!}({\bf p}^2+2a)^{-(1+2s)}\quad .
\eqno (3.22)
$$
Substituting this integral in Eq.~(3.21)
and integrating over momentum space gives
$$\ick{
V_{\rm eff}(\rho)={\lambda_1 \over 4}\rho^2 +a_3\rho^2
-{1 \over 64\pi}\lim_{s\to 0}({d \over ds})s{(2s)! \over s!s!}
&\rho^2\Bigl\{(-\lambda_1^2+{4\over \kappa^2}) \Bigr \}
{1 \over s}\Bigl ({\mu'^2 \over 2\lambda\rho}\Bigr)^{2s}.&(3.23) }
$$
Taking the derivative with respect to $s$
and the limit $s\rightarrow 0$, we obtain
$$\ick {
V_{\rm eff}(\rho)&={\lambda_1 \over 4}\rho^2
+{1\over 32\pi}(\lambda_1^2-{4\over\kappa^2})\rho^2\ln{2\lambda\rho\over\mu'^2}
+a_3 \rho^2
&(3.24) \cr }
$$
Note that the fermions did not contribute to the $\mu'$-dependent expression
of the scalar field effective
potential and their addition into the problem has no effect on the scalar
field effective potential.

Ultraviolet divergences are absent in this
method and renormalization is unnecessary.
The $a_3 \rho^2$-term  and
$\mu'$ dependences are eliminated to the expense of a new scale $\mu$
by normalizing the scalar field effective potential as in Eq.~(3.19).
The renormalized scalar field effective potential computed using OR in the
$R_\xi$-gauge in the $\xi\to 0$ limit is then given by Eq.~(3.20),
which agrees with the result given by the cutoff method in the same gauge
and with the scalar field effective potential computed with the use of both
methods in the Coulomb family gauges [see the appendix].

\medskip
\noindent {\bf IV. Scale Anomaly and Structure of the Effective Potential.}
\medskip

The advantage of Jackiw's method is the freedom in choosing the shifted field
to be independent of the classical equation of motion, at least in this model,
for the scalar field. It permits us to find the scalar field
effective potential as a
functional of the shifted field and look for a deformation of the scalar field
effective potential. We have computed the scalar field effective potential in
the $R_\xi$-gauge in the $\xi\to 0$ limit and in the appendix we perform the
calculation in the Coulomb family gauges for comparison and completeness, and
show that in this gauge fixing choice,
the scalar field effective potential is invariant with respect to
changes in the parameter $\xi$. We regulate the theory in both gauge fixings
using two regularization methods: a non-relativistic cutoff and operator
regularization. We find that the scalar field effective potential calculated in
either gauge fixing or either regularization methods is the same [see
Eqs.~(3.20) and (A.11)]. Furthermore, the scalar field effective
potential is transparent to the presence of a
constant magnetic field, {\it i.e.,} the scalar field effective potential does
not depend, to the order considered,
on a background gauge field, which satisfies
Gauss's law. As a spin-off of this result, we demonstrate that the results are
independent of the regularization scheme and hence prove indirectly that the
cutoff method regulates in a Galilean invariant way since OR preserves all
symmetries up to anomalies.

The first advantage of OR is that the method preserves all symmetries up
to anomalies. Another advantage
of OR is the absence of a mass $\mu'$-dependent term.
In the cutoff method, we show
the appearance of a cutoff
dependent mass term [see Eq.~(3.18) or Eq.~(A.10)]
necessitating the introduction of a counterterm, which
is not present at the classical level. Hence contrary to OR, the cutoff
method requires imposing a vanishing mass normalization condition.

Let us now analyse the scale anomaly.
Conformal symmetry is related to the
$\beta$-function. A non-vanishing $\beta$-function indicates
conformal symmetry breaking [14]. Using the renormalization group
equation
$$
0=\mu{d\over d\mu}V^R_{\rm eff}(\rho)=\Bigl [ \mu {\partial\over{\partial\mu}}
+\beta(\lambda_1(\mu)){\partial\over {\partial\lambda_1(\mu)}}
\Bigr ] V^R_{\rm eff}
(\rho) \eqno(4.1)
$$
the $\beta$-function reads
$$
\beta(\lambda_1(\mu))={1 \over 4\pi}\Bigl (\lambda^2_1(\mu)-{4 \over \kappa^2}
\Bigr )\quad .\eqno(4.2)
$$
For unrelated coupling constants the theory loses conformal symmetry.
At the self-dual point
$\lambda_1(\mu)=-{2 \over \kappa}$ and at $\lambda_1(\mu)={2\over \kappa}$
the
$\beta$-function vanishes; hence, the theory is conformally symmetric,
recovering the result of Lozano [12] and Bergman and Lozano [13].

We are now in a position to discuss the structure of the scalar field
effective potential of Eq.~(3.20)
following Coleman and Weinberg [15] since we have computed
radiative effects consistently with Gauss's law by including a constant
magnetic field on the plane.
In the present Chern-Simons theory, the coupling constant
$\lambda_1(\mu)$ can be either positive or negative in the range
$\lambda_1(\mu)>-{2\over \kappa}$. The theory with negative coupling constant
is well-defined since the Chern-Simons gauge field renders the
Hamiltonian positive definite. Consider first the simple case of the scalar
field theory of Eq.~(3.20) when ${1\over \kappa}\to 0$.
The theory with negative coupling constant is
undefined since the gauge field is absent. For the theory with positive
coupling constant, it seems that for small values of $\rho$ the
scalar field effective potential could
generate a minimum away from the origin. However, the location of this minimum
is outside the regime of validity since it is given by $\rho_{\rm min}=\mu \exp
(-{8\pi\over \lambda_1(\mu)})$. For small values of $\lambda_1(\mu)$,
$\rho_{\rm
min}$ is driven towards the origin and the scalar field
effective potential ceases to be
trustworthy.

In the case when the Chern-Simons term is present
a similar scenario arises when $\lambda_1(\mu) >0$ and
$\beta(\lambda_1(\mu)) >0$.
Then $\rho_{\rm min}=\mu^2\exp (-2{\lambda_1(\mu)\over \beta
(\lambda_1(\mu))})$
and this minimum is in the forbidden region.

An interesting scenario can arise when $\beta(\lambda_1(\mu))$ is negative.
The scalar field effective potential is the negative image of the usual
Mexican hat
potential. There are three possibilities.
When $\lambda_1(\mu)$ is positive with
$\lambda_1^2(\mu)\approx
{4\over \kappa^2}$, the location of the maximum is given by
\vfill\eject
{\noindent $\,$}$\rho_{\rm max}=\mu^2 \exp (2{ \lambda_1(\mu)
 \over |\beta(\lambda_1(\mu))|})$, which
may be driven too far away from the origin. The two other cases arise when the
Chern-Simons coupling diminishes such that $|\lambda_1(\mu)|\approx {1\over
2\pi\kappa^2}$. In this way, the radiative corrections are dominated by the
Chern-Simons contributions and are of the same order as the classical term in
the scalar field effective potential. For $\lambda_1(\mu)$
positive the maximum is located at $\rho_{\rm
max} \approx 7\mu^2$ and
for $\lambda_1(\mu)$ negative the maximum is located at $\rho_{\rm
max} \approx \mu^2$.
In the case where the classical potential is unbounded from below,
the region and an important distortion of the scalar field effective potential
around $\rho=0$. This feature could be important
when searching for topological vortex solutions.

In all three of these cases, when $\rho$ goes beyond the maximum, the scalar
field effective potential gets arbitrarily negative: it is unbounded from
below. One could expect higher-loop contributions to cure this behavior. In
any case, higher loop contributions would not perturb too much the scalar
field effective potential near the location of the maximum.

\vfill\eject

\medskip
\noindent {\bf Appendix. Effective Potential in the Coulomb Family Gauges}
\medskip

For sake of completeness, we compute the scalar field effective potential
in the Coulomb family gauges.
The authors of ref.[13] prove that in the $\xi\to 0$ limit of the
Coulomb family gauges, the background gauge field does not contribute
to the scattering of scalar-scalar fields, therefore in the functional
evaluation of the effective action, they should not contribute either to the
order considered in this paper.
In any case, for arbitrary $\xi$,
a similar proof as the one presented in section III can be used to show
that the background gauge field does not contribute to scalar field effective
potential to the order considered in the paper. Hence, we set
$a^\mu({\bf x})=0$ in this appendix. Also we ignore the fermion field, which
was seen to have no effect above.
We find then that the scalar field effective potential is $\xi$-independent.

We gauge fix this time the action (2.3) with the Galilean-invariant
Coulomb family gauges ${\cal L}_{G.F.}={1 \over 2\xi}(\nabla\cdot {\bf Q})^2$.
The quadratic action in the quantum fields
becomes
$$\ick
{\int dt d^2{\bf x}\; &\Bigl \{ {\kappa\over 2}(\partial_t{\bf Q})\times
{\bf Q}
- \kappa Q^0 {\bf \nabla}\times {\bf Q}
+ {1\over 2\xi}(\nabla\cdot{\bf Q})^2-{1\over 2}\rho{\bf Q}\cdot{\bf Q} \cr
& + i\pi^*\partial_t \pi- {1\over 2}\nabla\pi^*\nabla\pi
- {\lambda_1\over 4}
(\varphi^2 (\pi^*)^2+ 4 \varphi^*\varphi \pi^*\pi + (\varphi^*)^2(\pi)^2) \cr
&-\varphi(Q^0+{i\over 2}{\bf \nabla} \cdot {\bf Q})\pi^*
-\varphi^*(Q^0-{i\over 2}{\bf \nabla} \cdot {\bf Q})\pi
\Bigr \} &(A.0) \cr}
$$
hence
the quadratic part in the spacetime varying fields of the
shifted action (A.0) in the Coulomb family gauges has the form
$$\ick {\int dt d^2{\bf x}\;&{\cal L}'(\varphi, \pi^a(x),
Q^\mu(x))= \int dt dt' d^2{\bf x} d^2{\bf x'} \;\Bigl \{
{1\over 2} \pi^{*a}(x) {\cal D}^{-1}_{ab}(x-x') \pi^b(x')\cr
&-{1\over 2}Q^\mu(x) {\Delta}^{-1}_{\mu\nu}(x-x')Q^\nu (x')
\Bigr \}
+\int dt d^2{\bf x} \Bigl \{ J(x) \pi^*(x) + J^*(x)\pi(x) \Bigr \}
&(A.1) \cr}
$$
where the notation is the same as in section III, {\it i.e.,}
for the scalar field $\pi^a = (\pi, \pi^*)$;
$a=1,2$, $Q^\mu = ({Q^0\over c}, Q^1, Q^2)$
and $J(x)=~-\varphi(Q^0(x)+{i\over 2}\nabla\cdot{\bf Q(x)})$
with $J^*(x)$ its complex conjugate. However, we set the
coupling constant $\lambda_1=\lambda$.

The boson-gauge field transition induced by the last term is eliminated
by a conventional shift of variables, so that the quadratic part of the
resulting Lagrangian contains no such cross-terms.
The functional integral is then elementary.
The scalar field effective potential to ${\cal O}(\hbar)$ is then the sum
of all $n$-point one-loop graphs, and is given as
$$\ick {
V_{\rm eff}(\rho)={-\Gamma\over \int d^3x}=&\quad V_0(\rho)&(A.2) \cr
&- {i\over 2}
\int {d^2{\bf p}\over (2 \pi)^2}{d\omega\over 2\pi} \Bigl \{ \ln {\rm det}
{\cal D}^{-1}(\varphi;\omega,{\bf p^2}) + \ln {\rm det} \bigl (
{\Delta}^{-1}(\varphi;\omega,p^i)+N(\varphi;\omega,p^i)\bigr )
\Bigr \}
 }
$$
where again $V_0(\rho)={1 \over 2}\delta m\, \rho+{1 \over 4}(\lambda+\delta
\lambda)\rho^2$. $\delta m$ and $\delta\lambda$ are the counterterms,
anticipating the cutoff regularization method
necessary to render $V_{\rm eff}(\rho)$ finite.
The propagators in Fourier space are
$$
{\cal D}^{-1}(\varphi;\omega, p^i)
= \pmatrix { \omega -{1\over 2}{\bf p}^2-\lambda\rho
&-{\lambda\over 2}\varphi\varphi \cr
-{\lambda\over 2}\varphi^*\varphi^*
&-\omega-{1\over 2}{\bf p}^2-\lambda\rho\cr
}\quad ,\eqno (A.3)
$$

$$
\Delta^{-1}(\varphi;\omega, p^i)=
\pmatrix {0&-ic\kappa p^2 &ic\kappa p^1\cr
          ic\kappa p^2 &\rho-{1\over \xi}p^1p^1 &-i\kappa\omega-{1\over\xi}
p^1p^2\cr
          -ic\kappa p^1 &i\kappa\omega-{1\over \xi}p^1p^2 &
\rho-{1\over \xi}p^2p^2 \cr }\quad ,\eqno (A.4)
$$
and
$$
N(\varphi;\omega, p^i)={1\over {\rm det}{\cal D}^{-1}}
\pmatrix{ -c^2\rho({\bf p}^2 + \lambda\rho)
&c\rho\omega p^1 &c\rho\omega p^2 \cr
c\rho\omega p^1 & -{\rho\over 4}({\bf p}^2+3\lambda\rho)p^1p^1 &
-{\rho\over 4}({\bf p}^2+3\lambda\rho)p^1p^2 \cr
c\rho\omega p^2 &-{\rho\over 4}({\bf p}^2+3\lambda\rho)p^1p^2 &
-{\rho\over 4}({\bf p}^2+3\lambda\rho)p^2p^2 \cr},\eqno (A.5)
$$

\noindent {where} ${\rm det}{\cal D}^{-1}(\rho ;\omega, {\bf p^2})
=[-\omega^2 + {1\over 4}({\bf p}^2 +\lambda\rho)^2 +{\lambda\rho\over 2}
({\bf p}^2 +\lambda \rho)]$
and again $\varphi$
does not necessarily
satisfy the classical equation of motion.
A simple calculation of the determinants appearing in Eq.~(A.2)
shows that the first logarithm eliminates a term of the second logarithm;
after the cancellation, the remaining
part of the second logarithm reads
\vfill\eject
$$\ick {
V_{\rm eff}(\rho)=V_0(\rho)&-
{i\over 2}\int {d^2{\bf p}\over (2\pi)^2}{d\omega\over 2\pi}
\ln \Bigl [ {\bf p}^4 - \xi\lambda\rho^2\Bigr ] \cr
&-{i\over 2} \int {d^2{\bf p}\over (2\pi)^2} {d\omega\over 2\pi}
\ln \Bigl \{ -\omega^2 + ({{\bf p}^2\over 2})^2 +{1\over ({\bf p}^4
-\xi\lambda\rho^2)}\Bigl [ \lambda\rho {\bf p}^6 \cr
&\quad \quad \quad
+({3\over 4}\lambda^2 +{1\over \kappa^2})\rho^2{\bf p}^4
+{\lambda\over \kappa^2} \rho^3 {\bf p}^2 -{\xi\lambda\over \kappa^2}
(1+{3\over 4}\lambda^2)\rho^4  \Bigr ]\Bigr \}
&(A.6) \cr }
$$
We have dropped integrals that have no field dependence. Note that Eq.~(A.6)
is invariant under rotations and is a function of $\rho$ only.
Eq.~(A.6) contains all the necessary information to get the full
effective potential.

If one is interested in using the cutoff regularization
method, it is necessary to do the $\omega$ integration. We perform the
following change of variable $\tilde\omega =
\omega{\sqrt { {\bf p}^4-\xi\lambda\rho^2 }}$ and then make use of the
relation $-{i\over 2}\int {d\omega \over 2\pi} \ln \bigl \{
-\omega^2 + E^2 -i\epsilon \bigr \} = {1\over 2} E $
to write the effective potential in Coulomb family gauges as
$$\ick {
V_{\rm eff}(\rho)
= V_0(\rho)+{1\over 2}\int^{\Lambda^2} {d^2{\bf p}\over (2\pi)^2}
&{1\over {\sqrt {{\bf p}^4-\xi\lambda \rho^2}}} \Bigl [
{{\bf p}^8\over 4} +\lambda\rho {\bf p}^6 + ({3\over 4}\lambda^2 +
{1\over \kappa^2}-{\xi\lambda\over 4})\rho^2 {\bf p}^4 \cr
&+ {\lambda\rho^3\over \kappa^2}{\bf p}^2 - {\xi\lambda\over \kappa^2}
\rho^4 (1 + {3\over 4}\lambda^2) \Bigr ]^{1/2} &(A.7) \cr }
$$
The effective potential is infrared convergent since as ${\bf p^2}
\rightarrow 0$ the biggest contribution comes from the last term
$V_{\rm eff} - V_0 \approx {1\over 2} \int {d^2{\bf p}\over (2\pi)^2}
{1\over \sqrt { {\bf p}^4 -\xi \lambda\rho^2 } }
\Bigl ( -{\xi\lambda\over \kappa^2} \rho^4 (1+{3\over 4}
\lambda^2)\Bigr )^{1/2}$
which converges in the infrared sector for $\xi \neq 0$.
In the Landau gauge $\xi=0$, Eq.~(A.7) takes the simple form
$$\ick {
V_{\rm eff}(\rho)&=V_0(\rho)+{1\over 4}
\int^{\Lambda^2} { d^2{\bf p}\over (2\pi )^2} {1\over {\sqrt {\bf p^2}}}
{\sqrt {{\bf p}^2 +\lambda\rho}}{\sqrt { {\bf p}^2({\bf p}^2 + 3\lambda\rho)
+4{\rho^2\over \kappa^2} }}
&(A.8) \cr }
$$
which is again obviously infrared convergent since the element $d^2{\bf p}$
contain ${\sqrt{{\bf p}^2}}$ and therefore
cancels the denominator.

In the regime of small coupling constants $\lambda
<<1$ and $\kappa^{-1}<<1$, we can expand expression (A.7)
up to ${\cal O}(\rho^3)$ with $\xi\neq 0$ to obtain
\vfill\eject
$$\ick {
V_{\rm eff}(\rho)=&V_0(\rho)&(A.9)\cr
&+{1 \over 4}\int^{\Lambda^2} {d^2{\bf p} \over (2\pi)^2}
{\sqrt {{\bf p}^2 +\lambda\rho }} {\sqrt { {\bf p}^2 + 3\lambda\rho}}
\Bigl \{1+ {4\rho^2\over \kappa^2} {{\bf p}^2\over
({\bf p}^2+3\lambda\rho)({\bf p}^4 -\xi\lambda\rho^2)}
\Bigr \}^{1/2}
}$$
The evaluation of Eq.~(A.9) is then
straightforward and gives to ${\cal O}(\rho^3)$
$$\ick{
V_{\rm eff}(\rho)=V_0(\rho)+&
{1\over {32\pi}}\Bigl\{ 4\Lambda^2\lambda\rho + a_4\lambda^2\rho^2+
\bigl(-\lambda^2+{4\over \kappa^2}
\bigr)\rho^2\ln {\Lambda^2\over {\lambda\rho}}
 -2{\rho^2\over\kappa^2}\ln \bigl({-\xi\over \lambda}\bigr)\Bigr\}
&(A.10) \cr}
$$
where $a_4$ is a constant that could be evaluated if Eq.~(A.7) is solved
exactly. We do not compute it here since it does not contribute to the
renormalized effective
potential [see below]. Once the renormalization is performed with the
normalization conditions of Eq.~(3.19)
and vanishing mass, we find to ${\cal O}(\rho^3)$, in the cutoff method,
in the Coulomb family gauges
$$\ick{
V^R_{\rm eff}(\rho )=&{1 \over 4}\lambda_1(\mu)\rho^2
+{1 \over 32\pi}\bigl (\lambda_1(\mu)^2-{4 \over \kappa^2}
\bigr )\rho^2 \bigl (\ln{\rho \over \mu^2}-{3 \over 2}\bigr )\quad .&(A.11)
\cr }
$$
Notice that the renormalized scalar field
effective potential is $\xi$-independent and
agrees with Eq.~(3.20).

We now turn to evaluate the scalar field
effective potential using OR. We start with
the exact expression for $V_{\rm eff}(\rho)$ in Eq.~(A.6)
with the counterterms of $V_0(\rho)$ set to zero since there are
no ultraviolet-divergences in this method and hence no need for them.
We combine the two logarithms into one expression
and expand it to ${\cal O}(\rho^3)$
$$\ick {
V_{\rm eff}(\rho)={\lambda\over 4}\rho^2
-{i \over 2}\int {d^2{\bf p} \over (2\pi)^2} {d\omega \over 2\pi}\ln
{1\over \mu'^4}
&\Bigl \{ -\omega^2 + \bigl ({{\bf p}^2 \over 2} +\lambda\rho\bigr )^2
+ \bigl (
-{\lambda^2 \over 4}+ {1 \over \kappa^2} - {1 \over 2}\xi\lambda +
\xi\lambda {\omega^2 \over {\bf p}^4}\bigr )\rho^2 \cr
&+{\cal O}(\rho^3)
\Bigr \}. &(A.12) \cr  }
$$

Again,
the $n$-point function is easily extracted from the Schwinger expansion as
each $H_I$ corresponds to a 1-point insertion. In the case at hand,
$H_0={-\omega ^2+ \bigl ({{\bf p}^2 \over 2}+\lambda\rho\bigr )^2\over \mu'^4}$
and $H_I$ is the rest of the
expression in Eq.~(A.12). Now, we use Eqs.~(1.2-4) and get
$$\ick {
V_{\rm eff}(\rho)={\lambda \over 4}\rho^2
-{i\over 2}\lim_{s\to 0}\int {d^2 {\bf p}\over (2\pi)^2}
{d\omega\over 2\pi} &({d\over ds}){1\over \Gamma (s)}\times &(A.13)\cr
&\Bigl \{ \mu'^{4s}\rho^2 {
(-{1\over 4}\lambda^2 + {1\over\kappa^2}-{\xi\lambda\over 4}
+\xi\lambda{\omega^2\over {\bf p}^4})\over
[-\omega^2+({{\bf p}^2\over 2}+\lambda\rho)^2]^{1+s} } \Gamma (1+s)
+{\cal O}(\rho^3) \Bigr \} \cr }
$$
The integration over $\omega$ is again easy to perform in the complex plane
using the residue theorem.  The two integrals we need are the one in
Eq.~(3.22) and
$$\ick{
J\equiv
\int_{-\infty}^{\infty}
{d\omega\over 2\pi}{\omega^2\over \lbrack -\omega^2+({{\bf p}^2 \over
2}+a)^2\rbrack^{1+s}}=&i{(2s)!\over s!s!}{1\over 4}{1\over (1-2s)}({\bf p}^2+
2a)^{(1-2s)}\cr
&+(-)^{1+s}{R^{1-2s}\over (1-2s)}(1+e^{i2\pi s}) \quad .&(A.14) \cr }
$$
The quantity $R$ in $J$ is the radius of the curve enclosing the
singularity, which goes to infinity. We leave it there; however, it will
disappear from the final result.

	Substituting these integrals in Eq.~(A.13), cancelling
powers in ${\bf p}^2$,
and integrating over the momentum space, we obtain
$$\ick{
V_{\rm eff}(\rho)={\lambda \over 4}\rho^2 +a_5\rho^2
-{1 \over 64\pi}\lim_{s\to 0}({d \over ds})s{(2s)! \over s!s!}
&\rho^2\Bigl\{(-\lambda^2+{4\over \kappa}) +2\xi\lambda s\Bigr \}
{1 \over s}\Bigl ({\mu'^2 \over 2\lambda\rho}\Bigr)^{2s}.&(A.15) }
$$
The $R$-dependent expression coming from $J$ is convergent in the momentum
plane, and hence does not appear here because it is $\mu'$-independent. Again
the $a_5\rho^2$ term arises from the first term independent of $H_I$ in the
Schwinger expansion.
Taking the derivative with respect to $s$
and the limit $s\rightarrow 0$, we obtain
$$\ick {
V_{\rm eff}(\rho)&={\lambda \over 4}\rho^2
+{1\over 32\pi}(\lambda^2-{4\over\kappa^2})\rho^2\ln{2\lambda\rho\over\mu'^2}
+\Bigl ( {\xi\lambda\over 32\pi}+ a_5 \bigr ) \rho^2\quad .
&(A.16) \cr }
$$
Note that ultraviolet divergences are absent in this method and that
renormalization is unnecessary. The $\xi$ and $\mu'$ dependences are
eliminated by normalizing the scalar field effective potential as in
Eq.~(3.19). The result agrees with the result given by the cutoff method in
Eq.~(A.11) and with the result given in Eq.~(3.20) when the calculation is
performed in the $R_\xi$-gauge.

\vfill\eject
\centerline {\bf Acknowledgements}

	We thank G. Lozano, R.B. Mackenzie, and M.B. Paranjape for useful
comments.

\medskip

\centerline{\bf References}
\item{1.} J. Hong, Y. Kim and P.Y. Pac, Phys. Rev. Lett. {\bf 64}, 2230 (1990).
\item{2.} R. Jackiw and E.J. Weinberg, Phys. Rev. Lett. {\bf 64}, 2234 (1990).
\item{3.} See for a review, J. D. Lykken, J. Sonnenschein and N. Weiss,
Int. J. of Mod. Phys. {\bf A6}, 5155 (1991).
\item{4.} C. Lee, K. Lee and E.J. Weinberg, Phys. Lett. {\bf B243}, 105 (1990).
\item{5.} M. Leblanc and M.T. Thomaz, Phys. Lett. {\bf B281}, 259 (1992).
\item{6.} C.R. Hagen, Phys. Rev. {\bf D31}, 848 (1985).
\item{7.} R. Jackiw and S.Y. Pi, Phys. Rev. {\bf D49}, 3500 (1990).
\item{8.} G. Lozano, M. Leblanc and H. Min, Ann. of Phys.{\bf 219}, 328 (1992).
\item{9.} R. Jackiw, Phys. Rev. {\bf D9}, 1686 (1974).
\item{10.} D.G.C. McKeon, T.N. Sherry, Phys. Rev. {\bf 59}, 532 (1987); Phys.
Rev. {\bf D35}, 3584 (1987); Can. J. Phys. {\bf 66}, 268 (1988).
\item{11.} L. Culumovic, M. Leblanc, R.B. Mann, D.G.C. McKeon, and T.N. Sherry,
Phys. Rev. {\bf D41}, 514 (1990).
\item{12.} G. Lozano, Phys. Lett. {\bf B283}, 70 (1992).
\item{13.} O. Bergman and G. Lozano, MIT-CTP \# 2182 PREPRINT.
\item{14.} O. Bergman, Phys. Rev. {\bf D46}, 5474 (1992).
\item{15.} S. Coleman and E.J. Weinberg, Phys. Rev. {\bf D7}, 1888 (1973).
\end

\medskip
\centerline {\bf R\'ESUM\'E}
\medskip

Nous pr\'esentons le potentiel effectif pour un champ scalaire non relativiste
obtenu d'une th\'eorie de champs scalaire et fermionique auto-interagissants
coupl\'es \`a un champ de jauge dit de Chern-Simons. Le champ fermionique est
coupl\'e de mani\`ere non minimale au champ de jauge via une interaction du
type Pauli. La loi de Gauss relie lin\'eairement le champ magn\'etique aux
densit\'es de mati\`ere, obligeant ainsi l'inclusion d'un champ magn\'etique
classique uniforme. Nous montrons que le potentiel effectif n'est pas
affect\'e par
la pr\'esence de ce champ de jauge, \`a l'ordre consid\'er\'e. Nous calculons
ce potentiel dans deux jauges distinctes, soit dans une jauge $R_\xi$ o\`u
$\xi\to 0$, ainsi que dans la famille des jauges de Coulomb. Le potentiel
effectif du champ scalaire est le m\^eme quelque soit la fa\c con de fixer la
jauge et est ind\'ependant du param\`etre $\xi$ dans la famille des jauges de
Coulomb. La sym\'etrie conforme est bris\'ee  spontan\'ement sauf pour deux
valeurs de la constante de couplage des bosons, l'une d'entre elles \'etant
la constante de couplage autoduale. Au premier ordre perturbatif,
nous d\'emontrons que la structure du potentiel classique est chang\'ee
radicalement par les corrections radiatives et que le potentiel effectif
exhibe un minimum relatif autour de l'origine qui pourrait \^etre important
pour la d\'ecouverte de solutions dites ``vortex". Nous effectuons une
r\'egularisation de la th\'eorie avec les m\'ethodes ``operator regularization"
et ``cutoff" pour d\'emontrer que les r\'esultats sont ind\'ependants de la
mani\`ere de r\'egulariser.
\vfill \eject